\documentstyle[amssymb,preprint,aps]{revtex}
\begin{document}
\title{The anomaly of the oxygen bond-bending mode at 320 cm$^{-1}$ and the
additional absorption peak in the c-axis infrared conductivity of underdoped
YBa$_{2}$Cu$_{3}$O$_{7-\delta }$ single crystals revisited by ellipsometric
measurements}
\author{C. Bernhard$^{1}$, D. Munzar$^{1,2}$, A. Golnik$^{1\ast }$, C.T. Lin$^{1}$,
A. Wittlin$^{3}$, J. Huml{\'{i}}\v{c}ek$^{2}$, and M. Cardona$^{1}$}
\address{1) Max-Planck-Institut f\"{u}r Festk\"{o}rperforschung, Heisenbergstrasse 1,%
\\
D-70569 Stuttgart, Germany\\
2) Department of Solid State Physics and Laboratory of Thin Films and\\
Nanostructures, Faculty of Science, Masaryk University, Kotl\'{a}\v{r}sk\'{a}%
\\
2, CZ-61137 Brno, Czech Republic\\
3) Institute of Physics, Polish Academy of Sciences, Aleja L\'otnikow 32,\\
PL-02-668, Poland}
\date{\today}
\maketitle

\begin{abstract}
We have performed ellipsometric measurements of the far-infrared c-axis
dielectric response of underdoped YBa$_{2}$Cu$_{3}$O$_{7-\delta }$ single
crystals. Here we report a detailed analysis of the temperature-dependent
renormalization of the oxygen bending phonon mode at 320 cm$^{-1}$ and the
formation of the additional absorption peak around 400-500 cm$^{-1}$. For a
strongly underdoped YBa$_{2}$Cu$_{3}$O$_{6.5}$ crystal with T$_{c}$=52 K we
find that, in agreement with previous reports based on conventional
reflection measurements, the gradual onset of both features occurs well
above T$_{c}$ at T*$\sim $150 K. Contrary to some of these reports, however,
our data establish that the phonon anomaly and the formation of the
additional peak exhibit very pronounced and steep changes right at T$_{c}$.
For a less underdoped YBa$_{2}$Cu$_{3}$O$_{6.75}$ crystal with T$_{c}$=80 K,
the onset temperature of the phonon anomaly almost coincides with T$_{c}$.
Also in contrast to some previous reports, we find for both crystals that a
sizeable fraction of the spectral weight of the additional absorption peak
cannot be accounted for by the spectral-weight loss of the phonon modes but
instead arises from a redistribution of the electronic continuum. Our
ellipsometric data are consistent with a model where the bilayer cuprate
compounds are treated as a superlattice of intra- and inter-bilayer
Josephson-junctions.

\bigskip
\noindent PACS Numbers: 74.25.Gz, 74.25.Kc, 74.50.+r, 74.72.Bk
\end{abstract}

\allowbreak

\newpage

\section{Introduction}

It was early recognized that some of the infrared active c-axis phonon modes
of the high-T$_{c}$ cuprate superconductors (HTSC) exhibit rather strong
changes (so-called `phonon anomalies') in the vicinity of the
superconducting transition \cite{Litvinchuk1}. The most pronounced phonon
anomalies have been observed for those compounds which contain two (or
three) closely spaced CuO$_{2}$ layers per unit cell, like the bilayer
compounds YBa$_{2}$Cu$_{3}$O$_{7-\delta }$ (Y-123) \cite{Homes1,Schuetzmann1}%
, YBa$_{2}$Cu$_{4}$O$_{8}$ (Y-124) \cite{Basov1} and Pb$_{2}$Sr$_{2}$CaCu$%
_{2}$O$_{8}$ \cite{Reedyk1}, or the trilayer system Tl$_{2}$Ba$_{2}$Ca$_{2}$%
Cu$_{3}$O$_{10}$ \cite{Zetterer1}. In Y-123 and Y-124 the most pronounced
renormalization occurs for the so-called oxygen bond-bending mode at 320 cm$%
^{-1}$ which involves the in-phase vibration of the O(2) and O(3) oxygen
ions of the CuO$_{2}$ planes against the Y-ion which is located in the
center of the bilayer and against the ions of the CuO chains \cite{Henn1}.
The renormalization of the 320 cm$^{-1}$ mode is accompanied by the
formation of an additional broad absorption peak in the frequency range
between 400 to 500 cm$^{-1}$ at low temperature. The effects are most
spectacular for strongly underdoped YBa$_{2}$Cu$_{3}$O$_{6.5-6.6}$ with T$%
_{c}\sim $50-60 K. Here the phonon mode at 320 cm$^{-1}$ softens by almost
20 cm$^{-1}$ and loses most of its spectral weight which is transferred to
the additional broad peak. As the hole doping of the CuO$_{2}$ planes
increases, the anomaly of the 320 cm$^{-1}$ phonon mode becomes less
pronounced. Simultaneously, the additional peak shifts towards higher
frequencies and becomes considerably weaker.

Evidently, there exists an intimate relationship between the strong anomaly
of the oxygen bond-bending mode at 320 cm$^{-1}$ and the formation of the
additional absorption peak. The underlying mechanism, however, is yet
unknown and the subject of an ongoing discussion \cite
{Litvinchuk1,Hastreiter1,Hauff1}.

Recently van der Marel and coworkers have proposed a very interesting
explanation for the additional absorption peak \cite{vdMarel1,Grueninger1}.
They assumed that the CuO$_{2}$ planes of underdoped HTSC are not coherently
coupled, not even the closely spaced planes of the bilayers. From this point
of view a bilayer superconductor like Y-123 can be treated as a stack of
two-dimensional superconducting layers which forms a superlattice of intra-
and inter-bilayer Josephson-junctions. The dielectric response of such a
superlattice of Josephson junctions exhibits two zero-crossings
corresponding to two longitudinal Josephson-plasmons: the inter-bilayer and
the intra-bilayer one. In addition, it has a pole corresponding to the so
called `transverse optical Josephson plasmon' \cite{vdMarel1}. Van der Marel
have suggested that the additional absorption peak discussed above may
correspond to this transversal resonance. Very recently they have confirmed
their suggestion by more quantitative considerations regarding the doping
dependence of the peak position \cite{Grueninger1}. Some of us have shown
that this model can be extended to account not only for the presence of the
additional peak but also for the related phonon anomalies \cite{Munzar1}.
The essential idea consists in including the local electrical fields acting
on the ions that participate in the phonon modes. In particular, the model
has allowed us to explain the details of the anomaly of the 320 cm$^{-1}$
phonon mode in Y-123. This has been shown to arise from a dramatic change of
the local electrical field acting on the in-plane O(2)- and O(3)-ions caused
by the onset of inter- and intra-bilayer Josephson effects.

The verification of the existence of intra-bilayer Josephson plasmons, may
have rather far reaching consequences. The finding that even the
closely-spaced CuO$_{2}$ layers are only weakly (i.e. Josephson) coupled
along the c-axis would favor models which predict that the electronic ground
state of the CuO$_{2}$ planes is unconventional, involving charge
confinement to the planes and incoherent coupling between the planes in the
normal state \cite{Anderson1}. So far, the existence of the Josephson plasma
resonance has been firmly established only for the case of CuO$_{2}$ planes
(or pairs of planes) that are separated by insulating layers much wider than
the in-plane lattice constant. Such examples are the studies of the c-axis
transport \cite{Rapp1,Schlenga1}, the microwave absorption \cite{Matsuda1}
or the FIR c-axis conductivity \cite{Tamasaku1,Bentum1,Bentum2}. Even two
different longitudinal plasma modes have been recently observed in the T*
phase SmLa$_{1-x}$Sr$_{x}$CuO$_{4}$ which has two kinds of blocking layers:
the fluorite type Sm$_{2}$O$_{2}$ layers and the rocksalt-type (La,Sr)$_{2}$O%
$_{2}$ layers \cite{Shibata1}. \ 

In order to establish the above interpretation of the anomalies in the
c-axis conductivity of underdoped Y-123, two points have to be clarified
which have not been discussed in our previous communication \cite{Munzar1}.
Firstly, according to the model of the superlattice of inter- and
intra-bilayer Josephson junctions (lets call it a
Josephson-superlattice-model) the phonon anomalies should become very
pronounced only below T$_{c}$, while they can appear somewhat above T$_{c}$
for the strongly underdoped and most anisotropic samples as a result of
pairing fluctuations within the bilayers. From the previous experimental
results it was not clear whether this is the case. Reflectance measurements
performed on strongly underdoped Y-123 single crystals rather seemed to
indicate that the anomaly of the 320 cm$^{-1}$ phonon sets in well above T$%
_{c}$ at a temperature T*%
%TCIMACRO{\TEXTsymbol{>}}%
%BeginExpansion
\mbox{$>$}%
%EndExpansion
%TCIMACRO{\TEXTsymbol{>}}%
%BeginExpansion
\mbox{$>$}%
%EndExpansion
T$_{c}$ and proceeds without any noticeable change at T$_{c}$ \cite
{Schuetzmann1,Hauff1}. The temperature dependence of the phonon anomaly
rather seemed to resemble \cite{Schuetzmann1,Litvinchuk3} that of the
spin-lattice relaxation rate (T$_{1}$T)$^{-1}$ or the Knight shift observed
in NMR\ experiments, both of which are determined by the to the so-called
`spin-gap phenomenon', i.e., by a gradual and incomplete depletion of the
low energy spin excitations. These speculations are supported by\ Zn
substitution experiments: the Zn substitution is known to suppress the
spin-gap effect \cite{Kakurai1} and it has also been shown to remove the
anomaly of the 320 cm$^{-1}$ phonon mode and the additional absorption peak 
\cite{Hauff1}. Second, the model predicts that the additional peak should
acquire a considerable part of its spectral weight (SW) from the electronic
background, to be more specific from the superconducting condensate \cite
{Munzar1}. This prediction is not consistent with Ref. \cite{Homes1} where
it has been suggested that the SW of the additional peak is fully accounted
for by the spectral-weight loss ($\Delta $SW) of the phonon modes at 320 and
560 cm$^{-1}$.

In the following we report an ellipsometric study of the far-infrared (FIR)
c-axis dielectric response of two underdoped YBa$_{2}$Cu$_{3}$O$_{7-\delta }$
single crystal: one strongly underdoped ($\delta \approx $0.5, T$_{c}$=52
K), the other moderately underdoped ($\delta \approx $0.25, T$_{c}$=80 K).
In particular, we present a detailed analysis of the anomaly of the oxygen
bond-bending mode at 320 cm$^{-1}$ and of the additional absorption peak. In
case of the strongly underdoped crystal our ellipsometric measurements
establish that the temperature-evolution of the phonon anomaly and the
additional absorption peak exhibits a two-step behavior with a smooth onset
at T*$\sim $150K%
%TCIMACRO{\TEXTsymbol{>}}%
%BeginExpansion
\mbox{$>$}%
%EndExpansion
%TCIMACRO{\TEXTsymbol{>}}%
%BeginExpansion
\mbox{$>$}%
%EndExpansion
T$_{c}$=52 K followed by a sudden and steep change right at T$_{c}$=52 K.
For the moderately underdoped crystal the onset of the anomaly occurs in the
vicinity of T$_{c}$. For both crystals our data establish that the
absorption peak obtains only some part of its spectral weight from the
phonon system while a substantial part (at least 40-50\%) arises from the
electronic background.

\section{Experimental Technique}

\subsection{Sample preparation}

The YBa$_{2}$Cu$_{3}$O$_{7-\delta }$ single crystals with typical dimensions
of 2x2x(0.5-1) mm$^{3}$ have been grown in Y-stabilized Zr$_{2}$O crucibles 
\cite{Lin1}. For the ellipsometric measurements we used only crystals with a
smooth and shiny as-grown surface containing the c-axis. A strongly oxygen
deficient crystal has been prepared by sealing it in an evacuated quartz
tube together with a large amount of Y-123 powder whose oxygen content has
been previously adjusted to $\delta \approx $0.5 (by annealing in 0.1\% O$%
_{2}$ in Ar at 530$%
%TCIMACRO{\UNICODE[m]{0xb0}}%
%BeginExpansion
{{}^\circ}%
%EndExpansion
$ C and subsequently quenching into liquid nitrogen). The quartz-ampule
containing the crystal and the powder was annealed at 500 $%
%TCIMACRO{\UNICODE[m]{0xb0}}%
%BeginExpansion
{{}^\circ}%
%EndExpansion
$C for 10 days and subsequently slowly cooled to room temperature. A second,
moderately underdoped YBa$_{2}$Cu$_{3}$O$_{6.75}$ crystal has been prepared
by annealing in a flowing oxygen gas stream at 550 $%
%TCIMACRO{\UNICODE[m]{0xb0}}%
%BeginExpansion
{{}^\circ}%
%EndExpansion
$C and subsequent rapid quenching. The critical temperature T$_{c}$ and the
transition width $\Delta $T$_{c}$ (10 to 90\% of the diamagnetic shielding)
have been determined by DC-magnetization measurement in zero-field-cooled
(zfc) and field-cooled (fc) mode (H$_{ext}$=5 Oe) using a commercial SQUID
magnetometer.\ Figure 1 shows the temperature-dependent volume
suszeptibility, $\varkappa _{V}$, for the YBa$_{2}$Cu$_{3}$O$_{6.5}$ crystal
with T$_{c}$=52 K and $\Delta $T$_{c}$=3 K (open circles) and the YBa$_{2}$Cu%
$_{3}$O$_{6.75}$ crystal with T$_{c}$=80 K and $\Delta $T$_{c}$=4 K (solid
squares).

\subsection{\protect\bigskip\ Technique of far-infrared ellipsometry}

\bigskip The quantity measured in ellipsometry \cite{Azzam1} is the complex
reflectance ratio

$\widetilde{\rho }(\omega ,\Phi )=\widetilde{r}_{p}(\omega ,\Phi )/%
\widetilde{r}_{s}(\omega ,\Phi ),\qquad \qquad \qquad \qquad \qquad \qquad
\qquad \qquad \qquad \qquad \qquad (1)$

where $\Phi $ is the angle of incidence (in our experiment 80$%
%TCIMACRO{\UNICODE[m]{0xb0}}%
%BeginExpansion
{{}^\circ}%
%EndExpansion
$ with a beam divergence of $\pm $1.7$%
%TCIMACRO{\UNICODE[m]{0xb0}}%
%BeginExpansion
{{}^\circ}%
%EndExpansion
$) and $\widetilde{r}_{p}$ and $\widetilde{r}_{s}$ are the complex Fresnel
reflection coefficients for light which is polarized parallel ($p$) and
perpendicular ($s$) to the plane of incidence, respectively. The dielectric
function is extracted from $\widetilde{\rho }(\omega ,\Phi )$ by inverting
the Fresnel equations:

\bigskip $\widetilde{\epsilon }(\omega )=\left[ \left( 1-\widetilde{\rho }%
\left( \omega ,\Phi \right) \right) /\left( 1+\widetilde{\rho }\left( \omega
,\Phi \right) \right) \right] ^{2}\tan ^{2}\Phi \sin ^{2}\Phi +\sin ^{2}\Phi
.\qquad \qquad \qquad \qquad (2)$

This inversion assumes an isotropic sample. For an anisotropic sample, in
general, different elements of the dielectric tensor can contribute to $%
\widetilde{\rho }(\omega ,\Phi ).$ The formal inversion according to Eq. (2)
then yields only a so-called pseudodielectric function \cite{Aspnes1}. For
the case of YBa$_{2}$Cu$_{3}$O$_{7-\delta }$, which is an almost uniaxial
and strongly anisotropic material with metallic behavior of $\widetilde{%
\epsilon }_{a,b}(\omega )$ and insulating behavior of $\widetilde{\epsilon }%
_{c}(\omega )$, it was previously shown that the pseudodielectric function
represents a very good approximation for $\widetilde{\epsilon }_{c}$ when
the measurement is performed with the $c$-axis in the plane of incidence 
\cite{Henn2}.

The technique of ellipsometry provides significant advantages over
conventional reflection methods: i) it does not require the determination of
the absolute intensity of the reflected light (no reference problem) and ii)
the complex dielectric function $\widetilde{\varepsilon }=\epsilon
_{1}+i\epsilon _{2}$ is obtained directly, no Kramers- Kronig transformation
and thus no extrapolation of the reflectivity towards zero and infinite
frequency is needed \cite{Kircher1,Henn3}.

The ellipsometric measurements have been performed at the U4IR beamline of
the National Synchrotron Light Source (NSLS) at Brookhaven National
Laboratory (BNL), using a home-built setup attached to a Nicolet
Fast-Fourier Spectrometer \cite{Kircher1,Henn3}. The high brilliance of the
synchrotron light source enables us to perform very accurate ellipsometric
measurements in the far-infrared range even on samples with comparably small
ac-faces of 0.5$\times $1 mm$^{2}$. Since only relative intensities are
required, the ellipsometric measurements are very reproducible and the data
taken at a given temperature before and after thermal cycling or several
days of measurement coincide to within the noise level.

\section{Results}

\subsubsection{\protect\bigskip Strongly underdoped YBa$_{2}$Cu$_{3}$O$%
_{6.5} $}

Figure 2 shows the real part of the far-infrared (FIR) c-axis conductivity $%
\sigma _{c}$($\omega $,T) of the strongly underdoped YBa$_{2}$Cu$_{3}$O$%
_{6.5}$ crystal with T$_{c}$=52 K for (a) T=300, 200, and 150 K and (b)
T=150, 110, 60, 45, 35 and 4 K. At room temperature the six infrared-active
phonon modes at 155, 190, 280, 320, 560 and 630 cm$^{-1}$ are superimposed
on a weak and almost featureless electronic background. As the temperature
is lowered below room-temperature, the electronic background decreases
continuously and develops the so-called normal-state gap (NS gap) or
pseudogap which is a well known feature of the underdoped cuprate
superconductors \cite{Homes1,Tajima1,Bernhard1,Bernhard2,Bernhard3}.
Evidently, this NS-gap starts to develop at some rather high temperature, T$%
^{NG}\geq $ 300 K. Its characteristic frequency scale $\omega _{NG}$ is
fairly large and it even exceeds the measured spectral range, i.e., $\omega
_{NG}$%
%TCIMACRO{\TEXTsymbol{>}}%
%BeginExpansion
\mbox{$>$}%
%EndExpansion
700 cm$^{-1}$ \cite{Bernhard2,Bernhard3}. Figure 2 shows that the gradual
onset of the renormalization of the oxygen bond-bending mode at 320 cm$^{-1}$
and the additional broad peak at 410 cm$^{-1}$ occurs around T* $\sim $150
K, i.e., well below T$_{NG}\geq $300 K but also well above T$_{c}$=52 K.
However, the most important feature that is evident in Fig. 2b is that both,
the anomaly of the 320 cm$^{-1}$ phonon mode and the formation of the
additional absorption peak at 410 cm$^{-1},$ exhibit very pronounced and
steep changes right at T$_{c}$=52 K. This finding implies that both effects
are related to the superconducting transition rather than to the NS-gap or
to the spin-gap phenomenon as has been previously suggested \cite
{Schuetzmann1,Hauff1}. The gradual onset of the anomalies at T* $>$%
%TCIMACRO{\TEXTsymbol{>}}%
%BeginExpansion
\mbox{$>$}%
%EndExpansion
T$_{c}$ may be due to superconducting pairing fluctuations (especially
within the bilayers) which become particularly pronounced for strongly
underdoped and thus very anisotropic samples. We do not attempt here to
comment on the rather controversial question of whether the NS-gap and/or
the spin gap are also somehow related to the superconducting pairing
fluctuations.

We have performed a more quantitative data analysis by fitting to the
complex dielectric function in the spectral range 250%
%TCIMACRO{\TEXTsymbol{<}}%
%BeginExpansion
\mbox{$<$}%
%EndExpansion
$\omega $%
%TCIMACRO{\TEXTsymbol{<}}%
%BeginExpansion
\mbox{$<$}%
%EndExpansion
700 cm$^{-1}$ a sum of modified Lorentzian functions $\epsilon \left( \omega
\right) =S\cdot (\omega _{o}^{2}+i\Gamma \cdot Q)/\left( \left( \omega
_{o}^{2}-\omega ^{2}\right) -i\omega \cdot \Gamma \right) $ which represent
the contributions of the phonon modes at 280, 320, 560 and 630 cm$^{-1}$ $.$
This modified Lorentzian function which is obtained by mixing the real- and
the imaginary parts of the usual Lorentzian function (it basically
corresponds to a Fano-like function) allows one to perform Kramers-Kronig
consistent fits of asymmetric phonon lineshapes \cite{Homes1,Schuetzmann1}.
In agreement with previous reports \cite{Homes1,Schuetzmann1}, we find that
only the phonon modes at 560 and 630 cm$^{-1}$ are very asymmetric and that
the asymmetry Q of these modes increases with decreasing temperature. In
order to describe the flat electronic background and the additional broad
absorption peak in a Kramers-Kronig consistent way we have also included a
sum of seven broad Lorentzian oscillators which have been located between
250 and 700 cm$^{-1}$ and whose half-widths\ have been limited to values
between 150 and 500 cm$^{-1}$ .

In our opinion this fitting procedure does not allow one to obtain an
appropriate description of the phonon contribution since it assumes that all
the ions participating in the phonon modes experience the same (average)
electric field. Instead, as motivated in Ref. \cite{Munzar1}, we suggested
that the local electric fields acting on the ions exhibit significant
deviations from the average field as a result of the extremely weak
electronic coupling between the individual CuO$_{2}$ planes of underdoped
cuprates. We have shown that the spectacular anomaly of the 320 cm$^{-1}$
phonon mode, and also the asymmetry and the SW changes of the phonon modes
at 560 and 630 cm$^{-1},$ can be explained by such local field effects, in
particular, by the changes of the local electrical fields caused by the
onset of Josephson-effects within the intra- and inter-bilayer junctions 
\cite{Munzar1}. In the present paper we nevertheless apply the simpler
fitting procedure using modified Lorentzian functions in order to obtain the
temperature dependence of the parameters of the phonon modes which can be
readily compared with previous results.

Figure 3 shows the temperature dependence of (a) the oscillator strength S,
(b) the eigenfrequency $\omega _{o},$ and (c) the half-width $\Gamma $ of
the oxygen bond-bending mode at 320 cm$^{-1}$. It is evident from Fig. 3a
and 3b that the changes of S and $\omega _{o}$\ set in rather gradually
around T* $\sim $150 K. The temperature dependences of both quantities,
however, exhibit a sudden and steep change around T$_{c}$=52 K. This finding
contrasts with the previous reports that the renormalization of the 320 cm$%
^{-1}$ phonon mode does not exhibit any noticeable change around T$_{c}$ 
\cite{Schuetzmann1,Hauff1}. In agreement with the previous reports we find
that the half-width $\Gamma $ of the 320 cm$^{-1}$ mode starts to decrease
only below T$_{c}$=52 K \cite{Litvinchuk1,Homes1,Schuetzmann1,Litvinchuk2}.
For T*%
%TCIMACRO{\TEXTsymbol{>}}%
%BeginExpansion
\mbox{$>$}%
%EndExpansion
T%
%TCIMACRO{\TEXTsymbol{>}}%
%BeginExpansion
\mbox{$>$}%
%EndExpansion
T$_{c}$ it even tends to increase, but this small increase may be an
artifact of our fitting procedure. Figure 4 displays the electronic
background including the additional peak around 400 cm$^{-1}$ which has been
obtained by subtracting the contributions of the phonon modes at 280, 320,
560 cm$^{-1}$ and 630 cm$^{-1}$. The formation of the additional peak show
in Fig. 4 follows a similar temperature dependence like the anomaly of the
320 cm$^{-1}$ phonon mode. The broad peak gradually develops below T*$\sim $%
150 K and suddenly increases in magnitude below T$_{c}$=52 K. Evidently, the
peak position does not change much as a function of temperature. Below T$%
_{c} $ the peak is very pronounced and therefore hardly affected by the
subtraction of the phononic contribution. The frequency of the maximum
decreases slightly from 410 cm$^{-1}$ at 5 K to 395 cm$^{-1}$ at 45 K. Above
T$_{c}$ the peak becomes rather weak as compared to the phonon mode at 320 cm%
$^{-1}$. We therefore cannot reliably determine its position. Nevertheless
it is evident that it remains close to 400 cm$^{-1}$. Note that within the
Josephson-superlattice-model the peak position is expected to decrease only
by about 30 cm$^{-1}$ as the temperature is increased above T$_{c}$. The
position of the transverse plasmon is determined mainly by the intra-bilayer
Josephson-frequency which for strongly underdoped samples has been suggested 
\cite{Munzar1} not to exhibit any pronounced changes at T$_{c}$ since the
pairing fluctuations within the individual bilayers persist far above the
macroscopic critical temperature T$_{c}$.

Figure 5 shows a comparison of the temperature dependence of the spectral
weight of the additional 410 cm$^{-1}$ peak, SW$^{410}$, (solid squares)
with the spectral weight loss of the 320 cm$^{-1}$ phonon mode, $\Delta $SW$%
^{320}$, (open circles). The value of $\Delta $SW$^{320}$ has been
calculated according to the formula $\Delta $SW$=\pi ^{2}c\cdot \epsilon
_{o}\cdot \left[ S\left( 175K\right) -S\left( T\right) \right] \cdot \nu
_{o}^{2}$ using the oscillator strenghts S(T) and the eigenfrequencies $\nu
_{o}$ given in Fig. 3a and 3b. The error bar indicates the upper limit for
the spectral weight loss of the apical oxygen mode at 560 cm$^{-1}$. The
inset of Fig. 5 illustrates how we have estimated the spectral weight of the
additional peak SW$^{410}$ (shaded area) by integrating the conductivity
between 220 and 680 cm$^{-1}$ (open circles) and subtracting the
contribution of a linear electronic background (solid line). Note that our
assumption of a linear electronic background conductivity is supported by
the high temperature data for T%
%TCIMACRO{\TEXTsymbol{>}}%
%BeginExpansion
\mbox{$>$}%
%EndExpansion
T*=150 K (see Fig. 4a) and by the data for less strongly underdoped samples
where the additional peak is comparably weak and appears only below T$_{c}$
(see \ Ref. \cite{Bernhard2,Bernhard3} and discussion below). It is evident
from Fig. 5 that the spectral weight losses of the phonon modes at 320 and
560 cm$^{-1}$ account only for some fraction of the spectral weight of the
additional absorption peak at 410 cm$^{-1}$. We estimate that at least 40-50
\% of the spectral weight of the additional absorption peak does not arise
from the phonon subsystem but instead seems to arise from the electronic
continuum. Such a redistribution of the electronic spectral weight towards
the additional absorption peak is predicted by the
Josephson-superlattice-model where the transverse Josephson plasmon acquires
a sizeable fraction of the spectral weight of the SC condensate \cite
{Grueninger1,Munzar1}. This means that some fraction of the spectral weight
removed in the FIR-regime does not appear in the delta-function at zero
frequency which describes the inductive response of the superfluid
condensate but is instead shifted to the additional absorption peak
representing the transversal Josephson plasmon. In fact, our choice of the
linear electronic background (solid line) is rather conservative. Also shown
in the inset is a fit with a Lorentzian plus a linear electronic background
conductivity (dashed lines) which gives a somewhat larger value of SW$%
^{410}\approx $ 16.000 $\Omega ^{-1}$cm$^{-2}$ at T=5 K. The result of this
fit agrees surprisingly well with the prediction of the
Josephson-superlattice model that the SW of the bare transverse
Josephson-plasmon (neglecting the interaction with the phonons and the
consequent redistribution of the spectral weight of the 320 cm$^{-1}$ phonon
mode, $\Delta $SW$^{320}(5K)\approx 5000$ $\Omega ^{-1}$cm$^{-2}$)\ should
be around 10.000 $\Omega ^{-1}$cm$^{-2}$ \cite{Munzar1}.

It has been reported previously that the spectral weight of the additional
absorption peak can be fully accounted for by the spectral weight loss of
the phonon modes at 320 and 560 cm$^{-1}$ \cite{Homes1}. We note that this
discrepancy between the conclusions of \cite{Homes1} and ours does not arise
from any significant differences in the experimental data but rather
originates from the difference in the estimate of the electronic background.
It seems that Homes et al. obtained a lower spectral weight of the
additional absorption peak by introducing an electronic background which
develops a gap-like feature around 280 cm$^{-1}$ (see for example Fig. 10 of
Ref. \cite{Homes1}). They argued that this gap feature is related to the NS
gap. Our recent ellipsometric data, however, do not support such an
interpretation because they show that the onset frequency of the NS gap, $%
\omega _{NG}$, is significantly larger: it increases from $\omega
_{NG}\approx $650 cm$^{-1}$ for slightly underdoped samples to $\omega
_{NG}>>700$ cm$^{-1}$ for strongly underdoped samples. We have also shown
that the temperature at which the signatures of the normal state NS-gap
start to appear increases rather rapidly on the underdoped side and even
exceeds room temperature for strongly underdoped samples\ \cite
{Bernhard2,Bernhard3} (see also Fig.2). Note that recent ARPES- and
tunneling experiments yield a similar temperature- and doping dependence of
the NS-gap (even the absolute values of the gap size agree reasonably well) 
\cite{Ding1,Loeser1,Norman1,Miyakawa1}. This fact allows us to conclude that
the characteristic energy- and temperature scales of the additional
absorption peak and the NS-gap (and also their doping dependence) are very
different which makes it rather unlikely that both have a common origin.

\subsubsection{Moderately underdoped YBa$_{2}$Cu$_{3}$O$_{6.75}$}

We have also performed FIR ellipsometric measurements on a less strongly
underdoped YBa$_{2}$Cu$_{3}$O$_{6.75}$ sample with T$_{c}$=80 K. Figure 6
shows its c-axis conductivity $\sigma _{c}$($\omega $,T) at different
temperatures between 300 and 10 K. The anomaly of the oxygen bond-bending
mode at 320 cm$^{-1}$ is significantly weaker than that of the strongly
underdoped sample. The additional absorption peak is located at higher
frequencies (around $480$ cm$^{-1}$) and it is less pronounced. Figure 7
shows the temperature dependence of (a) the oscillator strength S, (b) the
eigenfrequency $\omega _{o}$ and (c) the half-width $\Gamma $ of the 320 cm$%
^{-1}$ phonon mode which have been obtained using the fitting procedure
outlined above. It can be seen that the onset of the anomaly of the 320 cm$%
^{-1}$ mode occurs now close to the superconducting transition (i.e., T* $%
\lesssim $ 100 K). A similar result has been previously obtained from
conventional reflection measurements on weakly underdoped Y-123 crystals 
\cite{Schuetzmann1}. The formation of the NS-gap, as evidenced from the
gradual suppression of the electronic background conductivity, can be seen
to set in at a significantly higher temperature T$_{NG}\gtrsim $200 K and
the characteristic frequency scale of the NS-gap still exceeds the measured
spectral range, $\omega _{NG}$%
%TCIMACRO{\TEXTsymbol{>}}%
%BeginExpansion
\mbox{$>$}%
%EndExpansion
700 cm$^{-1}$. These findings supports our point of view that the anomaly of
the 320 cm$^{-1}$ mode and the formation of the additional absorption peak
are not directly related to the mechanism that is responsible for the NS-gap 
\cite{Bernhard2,Bernhard3}. Figure 8 shows the electronic background
including the additional absorption peak which has been obtained by
subtracting the contributions of the phonon modes at 280, 320, 570 cm$^{-1}$
and 620 cm$^{-1}$ (as described above). Figure 9 shows the estimated
T-dependences of the spectral weight of the additional peak at 480 cm$^{-1}$
(once more assuming a linear electronic background) and of the spectral
weight loss ($\Delta $SW) of the phonon mode at 320 as estimated from the
change in its oscillator strength according to: $\Delta $SW$=\pi ^{2}c\cdot
\epsilon _{o}\cdot \left[ S\left( 120K\right) -S\left( T\right) \right]
\cdot \nu _{o}^{2}$. In analogy to the strongly underdoped sample, a
sizeable fraction of the spectral weight of the absorption peak does not
seem to arise from the phonon subsystem.

\subsection{\protect\bigskip Summary}

In summary, we have performed ellipsometric measurements of the FIR c-axis
conductivity of underdoped YBa$_{2}$Cu$_{3}$O$_{7-\delta }$ single crystals.
In particular, we have studied the temperature dependence of the spectacular
anomaly of the oxygen bond-bending mode at 320 cm$^{-1}$ and of the
additional absorption peak. For the strongly underdoped YBa$_{2}$Cu$_{3}$O$%
_{6.5}$ crystal with T$_{c}$=52 K the gradual onset of the anomaly of the
320 cm$^{-1}$ phonon mode and the growth of the additional absorption peak
around 400 cm$^{-1}$ occur around T*$\sim $150 K, i.e., well below the onset
temperature of the NS-gap, T$_{NG}\gtrsim $300 K, but also well above T$_{c}$%
=52 K. Most remarkably, however, we find that both anomalies exhibit a very
pronounced and marked change right at T$_{c}$=52 K. For a less strongly
underdoped YBa$_{2}$Cu$_{3}$O$_{6.75}$ crystal with T$_{c}$=80 K both
anomalies start to appear in the vicinity of the superconducting transition,
T*$\sim $T$_{c}$=80 K while the signatures of the NS-gap appear already
below T$_{NG}\gtrsim $200 K. Our measurements thus establish that both
anomalies are related to the superconducting transition. For strongly
underdoped and thus very anisotropic samples, the gradual onset of the
anomalies may occur well above T$_{c}$ due to superconducting pairing
fluctuations within the individual bilayers, but there is always a steep and
sudden increase of both anomalies when the macroscopically coherent
superconducting state forms at T$_{c}$. This implies that the anomalies are
not related to the NS-gap in the c-axis conductivity nor to the spin-gap
phenomenon as observed in NMR- and NQR-measurements both of which exhibit no
noticeable changes at T$_{c}$. For both underdoped crystals a sum-rule
analysis of the changes of the spectral weights indicates that the spectral
weight of the additional absorption peak is not fully accounted for by the
spectral weight loss of the phonon modes at 320 and 560 cm$^{-1}$. At least
40-50 \% of the spectral weight of the additional absorption peak seems to
arise from the electronic background, namely from the superconducting
condensate. We have outlined that all the reported features are compatible
with a recently proposed model where the bilayer cuprate compounds such as
Y-123 are treated as a superlattice of inter- and intra-bilayer
Josephson-junctions. The additional absorption peak can be related to the
transverse optical plasmon, while the spectacular phonon anomaly can be
explained as due to the drastic changes of the local electrical fields
acting on the in-plane oxygen ions as the Josephson current sets in below T$%
_{c}$ for the inter-bilayer junctions while below T*$\geq $T$_{c}$ for the
intra-bilayer junctions.

\section{Acknowledgement:}

We gratefully acknowledge the support of G.P. Williams and L. Carr at the
U4IR beamline at NSLS. We also thank D. B\"{o}hme and W. K\"{o}nig for
technical help and E. Br\"{u}cher and R.K. Kremer for performing the SQUID
measurements. D.M. gratefully acknowledges support by the Alexander von
Humboldt Foundation.

\section{Figure Captions}

\bigskip

Figure 1: Temperature dependence of the zero-field-cooled (zfc) and
field-cooled (fc) volume DC-magnetization, $\chi _{V}$, of the YBa$_{2}$Cu$%
_{3}$O$_{6.5}$ crystal (open circles) and the YBa$_{2}$Cu$_{3}$O$_{6.75}$
crystal (solid squares). The external field H$_{ext}$=5 Oe was applied along
the c-axis of the platelet shaped crystals. Corrections for demagnetization
factors have not been taken into account. \ The critical temperatures and
the transition width (between 10 and 90\% of the diamagnetic shielding) are T%
$_{c}$=52 K, $\Delta $T$_{c}$= 3 K and T$_{c}$=80 K, $\Delta $T$_{c}$=4 K.

\bigskip

Figure 2: Temperature dependence of the real part of the FIR c-axis
conductivity of the strongly underdoped YBa$_{2}$Cu$_{3}$O$_{6.5}$ crystal
with T$_{c}$=52 K. Spectra are shown for (a) T=300, 200 and 150 K and (b)
T=150, 60, 45, 35 and 5 K.

\bigskip

Figure 3: Temperature dependence of (a) the oscillator strength S, (b) the
eigenfrequency $\omega _{o}$ and (c) the half width $\Gamma $ of the oxygen
bond-bending mode of the strongly underdoped YBa$_{2}$Cu$_{3}$O$_{6.5}$
crystal (T$_{c}$=52 K). The phonon parameters have been obtained by fitting
modified Lorentzian functions to the complex dielectric function (as
outlined in the text). The dashed line marks the superconducting transition
temperature T$_{c}$=52 K, the dotted line indicates the gradual onset of the
renormalization of the phonon mode around T*$\sim 150K$.

\bigskip

Figure 4: Temperature dependence of the electronic conductivity including
the additional absorption peak at 410 cm$^{-1}$ (presumed to be also of
electronic origin) of the strongly underdoped YBa$_{2}$Cu$_{3}$O$_{6.5}$
crystal with T$_{c}$=52 K which has been obtained by subtracting the
contributions of the phonon modes at 280, 320, 560 and 630 cm$^{-1}$.

\bigskip

Figure 5: \ Temperature dependence of the spectral weight of the additional
absorption peak at 410 cm$^{-1}$ (SW$^{410}$, solid squares) and the
spectral weight loss of the oxygen bond-bending mode at 320 cm$^{-1}$ ($%
\Delta $SW$^{320}$, open circles). The inset illustrates how the SW of the
additional absorption peak has been obtained. The open circles represent the
spectrum at T=5 K obtained after the phonons have been subtracted, the solid
line represent the linear background which has been further subtracted in
order to obtain SW$^{410}$ as indicated by the shaded area. Also shown is a
fit with a Lorentzian plus a linear background (dashed lines) which gives a
somewhat larger value of SW$^{410}\approx 16.000$ $\Omega ^{-1}cm^{-2}$ at 5
K. The spectral weight loss of the 320 cm$^{-1}$ phonon mode has been
calculated according to the formula $\Delta $SW$^{320}=\pi ^{2}c\cdot
\epsilon _{o}\cdot \left[ S\left( 175K\right) -S\left( T\right) \right]
\cdot \nu _{o}^{2}$ using the oscillator strenght S(T) and the
eigenfrequencies $\nu _{o}$ given in Fig. 3a and 3b. The error bar indicates
the maximum spectral weight loss at T= 5K of the phonon mode at 560 cm$^{-1}$%
.

\bigskip

Figure 6: Temperature dependence of the real part of the FIR c-axis
conductivity of the weakly underdoped YBa$_{2}$Cu$_{3}$O$_{6.75}$ crystal
with T$_{c}$=80 K.

\bigskip

Figure 7: Temperature dependence of (a) the oscillator strength S, (b) the
eigenfrequency $\omega _{o}$ and (c) the half width $\Gamma $ of the oxygen
bond-bending mode of the moderately underdoped YBa$_{2}$Cu$_{3}$O$_{6.75}$
crystal (T$_{c}$=80 K). The phonon parameters have been obtained by fitting
modified Lorentzian functions to the complex dielectric function (as
outlined in the text). The dashed line marks the superconducting transition
temperature T$_{c}$=80 K.

\bigskip

Figure 8: Temperature dependence of the electronic conductivity, including
the additional absorption peak around 480 cm$^{-1},$ of the moderately
underdoped YBa$_{2}$Cu$_{3}$O$_{6.75}$ crystal with T$_{c}$=80 K which has
been obtained by subtracting the contributions of the phonon modes at 280,
320, 570 and 620 cm$^{-1}$.

\bigskip

Figure 9: Temperature dependence of the spectral weight of the additional
absorption peak at 480 cm$^{-1}$ (SW$^{480}$, solid squares) and the
spectral weight loss of the oxygen bond-bending mode at 320 cm$^{-1}$ ($%
\Delta $SW$^{320}$, open circles). The spectral weight loss of the 320 cm$%
^{-1}$ phonon mode has been calculated according to the formula $\Delta $SW$%
=\pi ^{2}c\cdot \epsilon _{o}\cdot \left[ S\left( 175K\right) -S\left(
T\right) \right] \cdot \nu _{o}^{2},$ using the oscillator strenghts S(T)
and the eigenfrequencies $\nu _{o}$ given in Fig. 7a and 7b. The error bar
indicates the maximum spectral weight loss at T= 5K of the phonon mode at
570 cm$^{-1}$.

\bigskip\ 

\bigskip

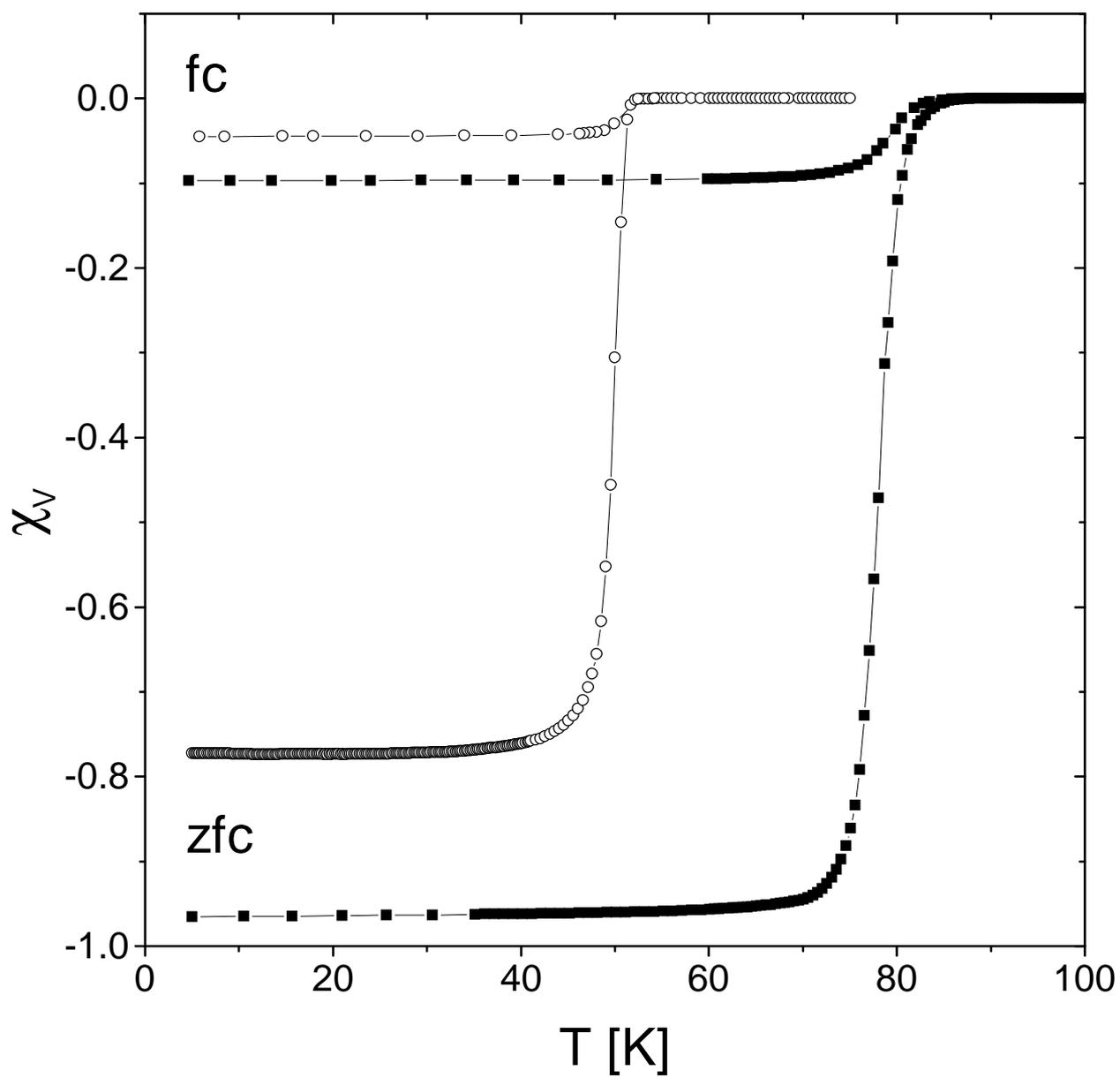
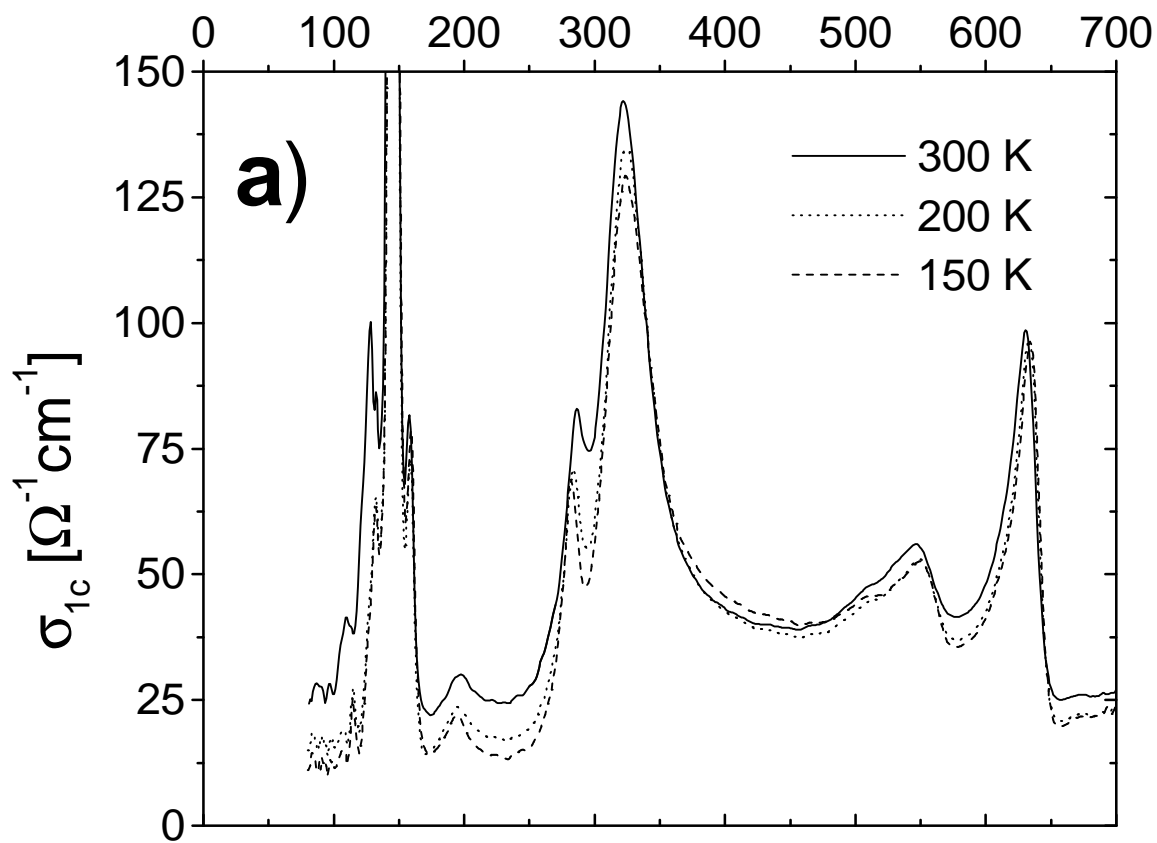
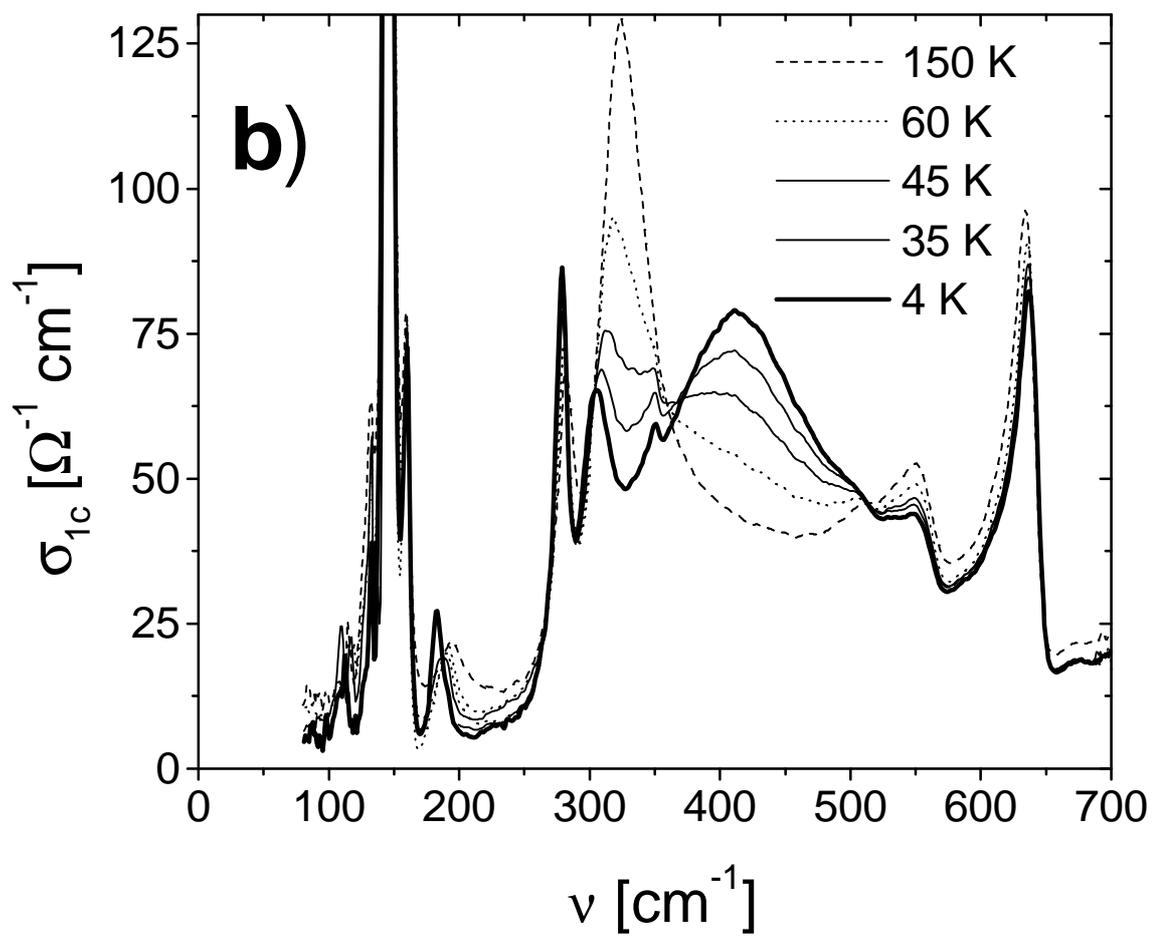
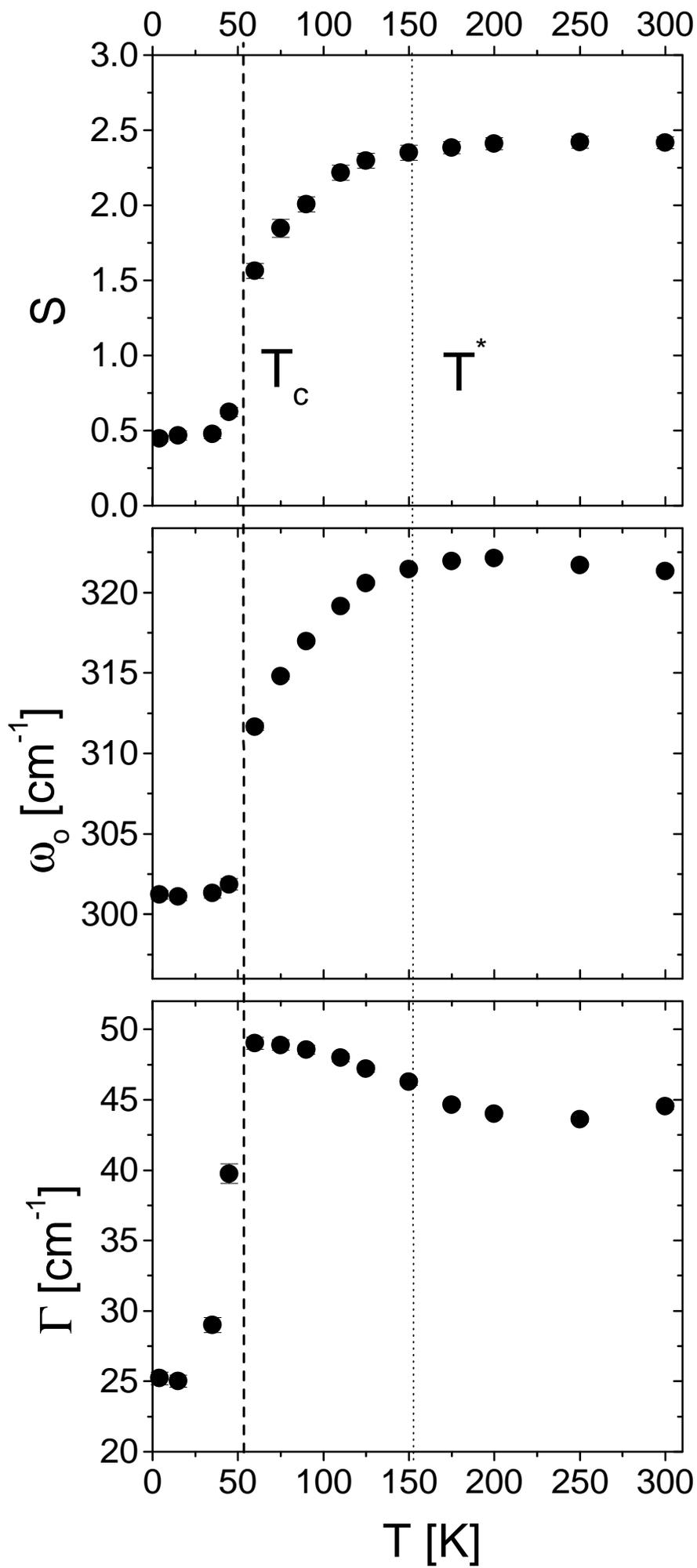
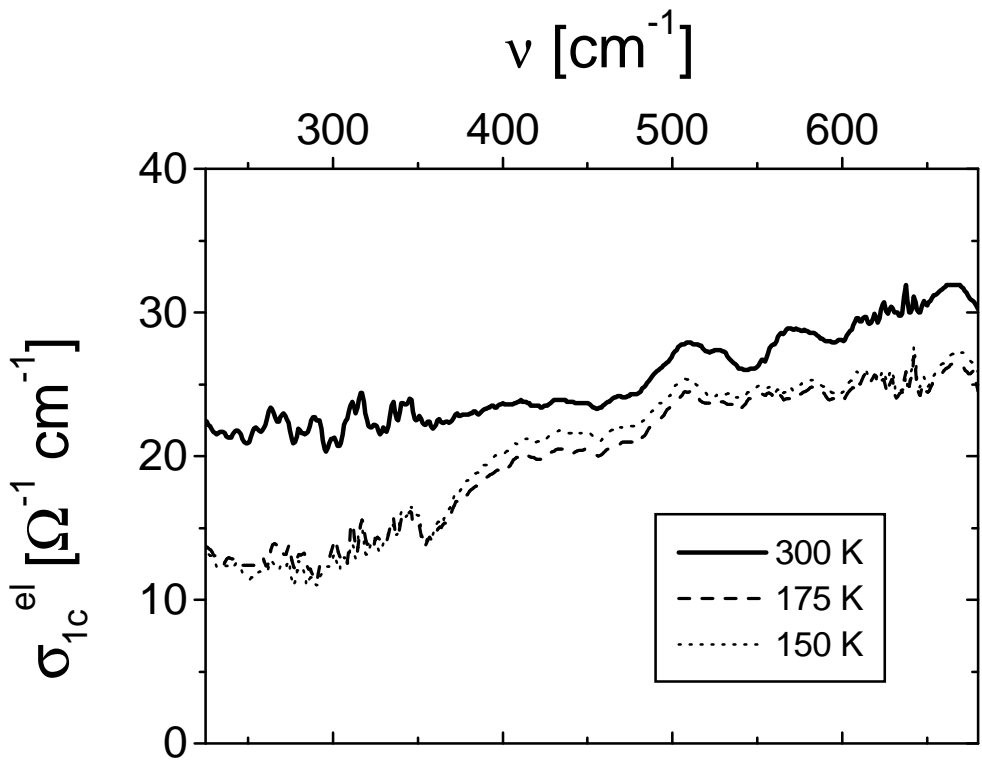
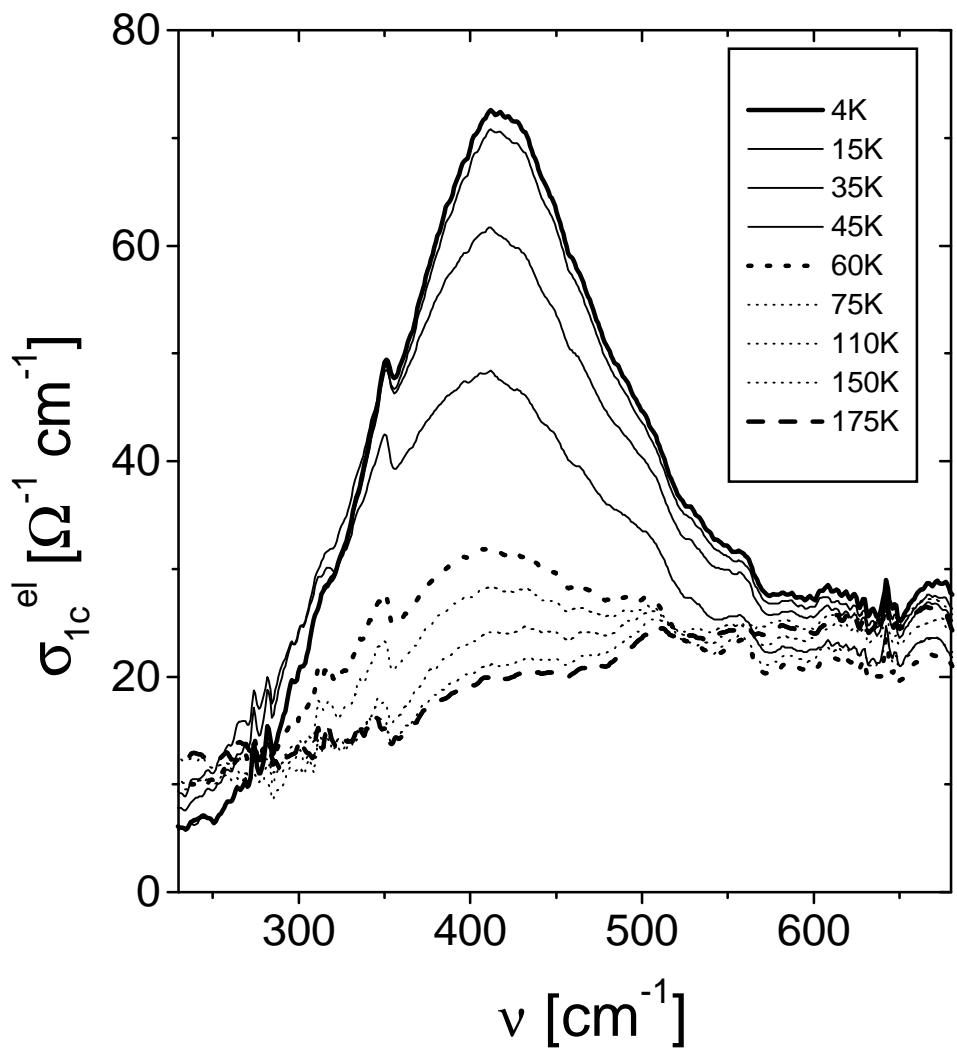
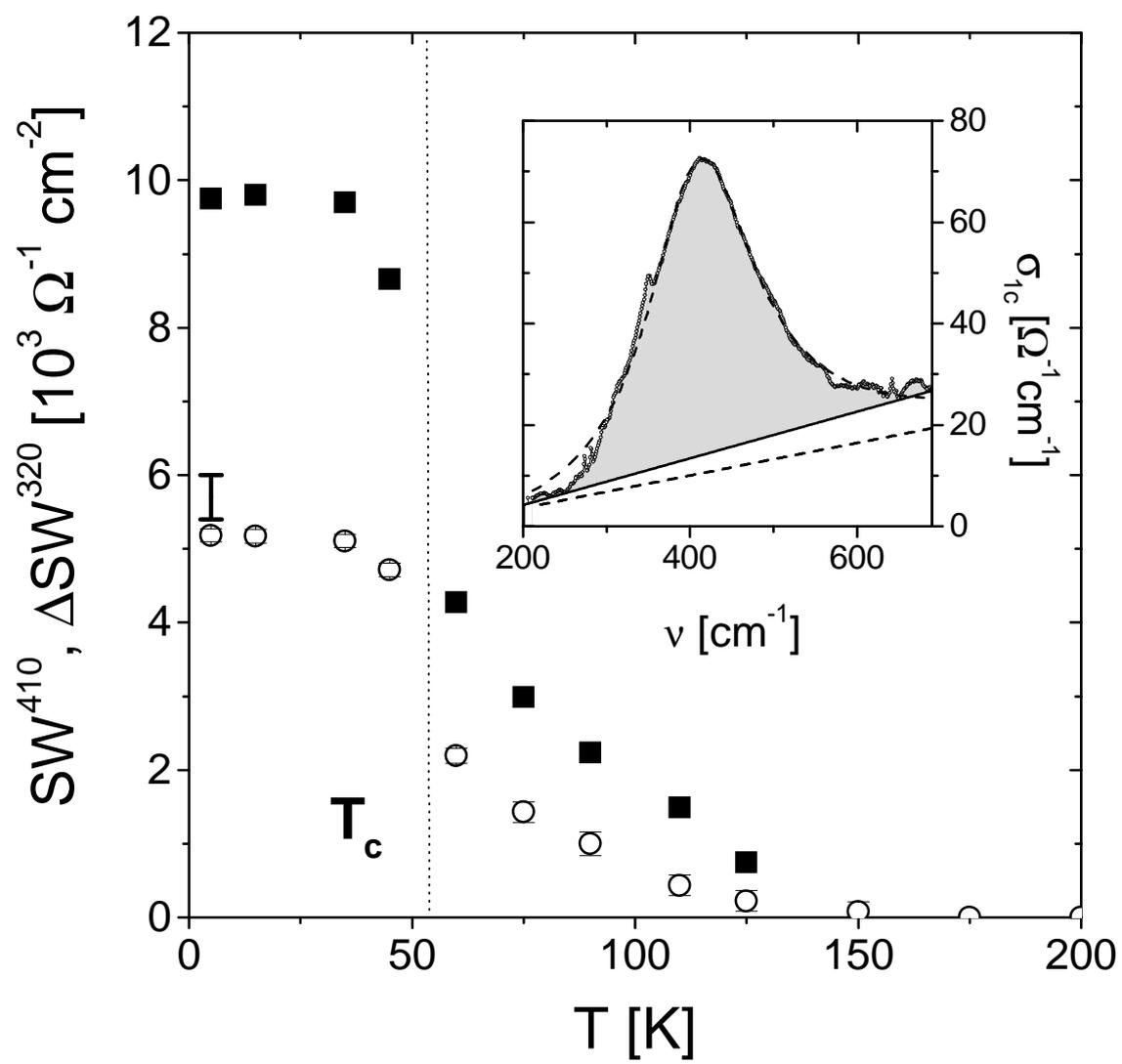
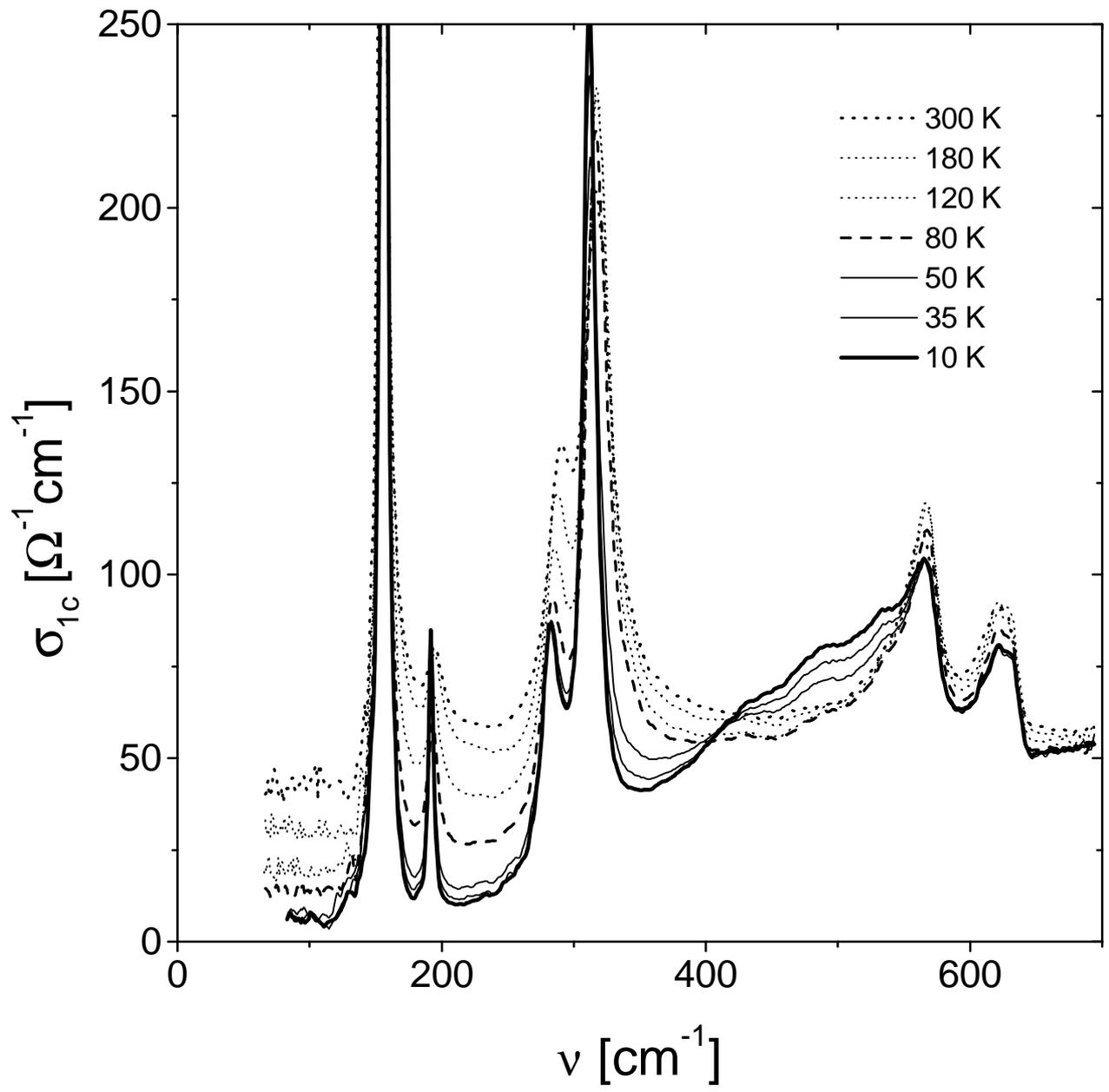
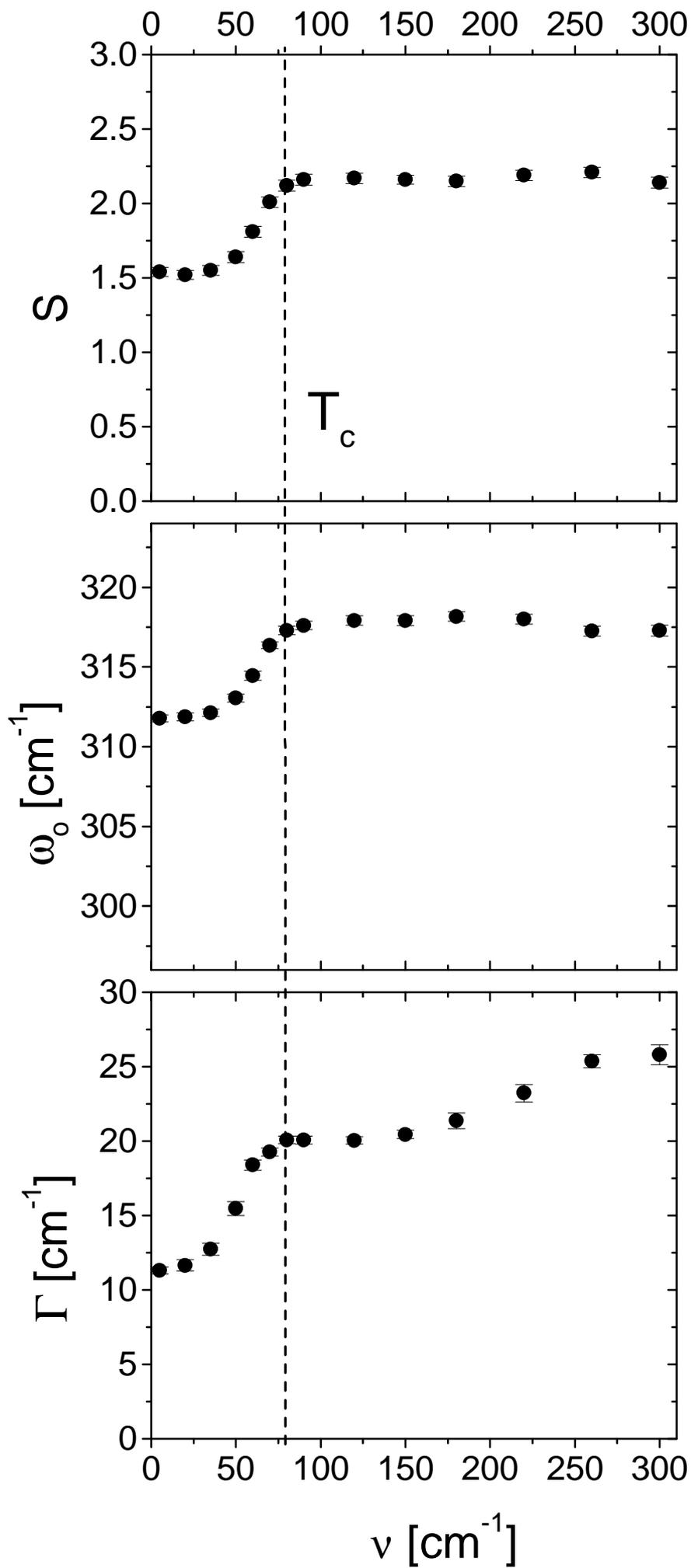
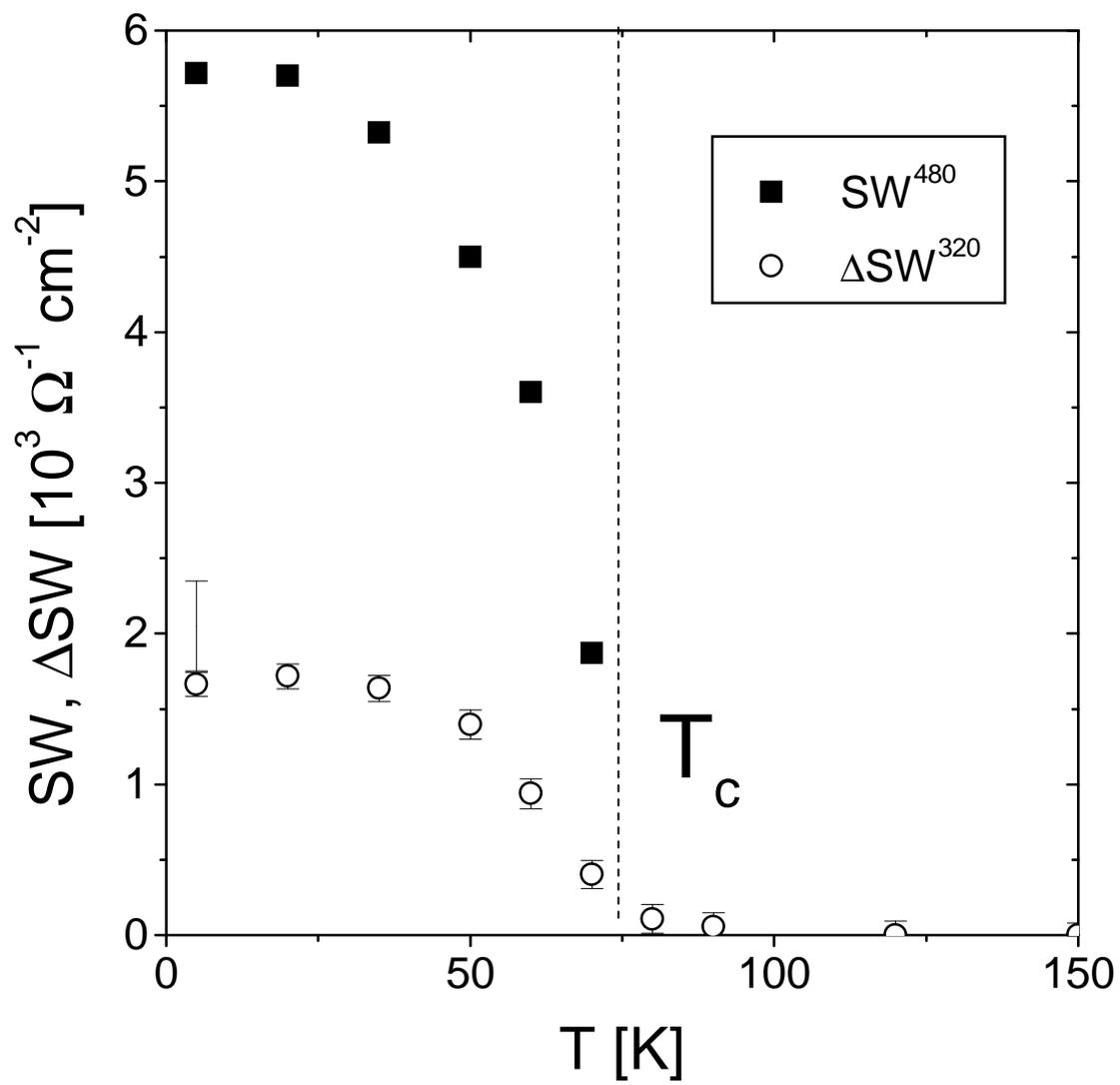
\end{document}